\documentclass[aps,prd,showpacs]{revtex4}
\usepackage{epsf,epsfig,graphicx,axodraw}
\begin{document}

\begin{flushright}
SI-HEP-2010-19 \\
\end{flushright}

\title{Next-to-leading-order correction to pion form factor in $k_T$ factorization}

\author{Hsiang-nan Li$^{a,b,c}$, Yue-Long Shen$^{a,d}$,  Yu-Ming Wang$^{e,f}$ , Hao Zou$^f$ }
\affiliation{{\it \small $^a$Institute of Physics, Academia Sinica, Taipei, Taiwan 115, Republic of China }\\
{\it \small $^b$Department of Physics, National Cheng-Kung University, Tainan, Taiwan 701, Republic of China}\\
{\it \small $^c$Department of Physics, National Tsing-Hua
University, Hsinchu, Taiwan 300, Republic of China}\\
{\it \small $^d$College of Information Science and Engineering,
Ocean University of China, Qingdao, Shandong 266100, P.R. China} \\
{\it \small $^e$Theoretische Physik 1, Fachbereich Physik,
Universit\"at Siegen, D-57068 Siegen, Germany}\\
{\it \small $^f$Institute of High Energy Physics and Theoretical
Physics Center \\ for Science Facilities, P.O. Box 918(4) Beijing
100049,  China}}

\begin{abstract}

We calculate next-to-leading-order (NLO) correction to the pion
electromagnetic form factor at leading twist in the $k_T$
factorization theorem. Partons off-shell by $k_T^2$ are considered
in both quark diagrams and effective diagrams for the
transverse-momentum-dependent (TMD) pion wave function. The
light-cone singularities in the TMD pion wave function are
regularized by rotating the Wilson lines away from the light cone.
The soft divergences from gluon exchanges among initial- and
final-state partons cancel exactly. We derive the infrared-finite
$k_T$-dependent NLO hard kernel for the pion electromagnetic form factor
by taking the difference of the above two sets of diagrams. Varying
the renormalization and factorization scales, we find that the NLO
correction is smaller, when both the scales are set to the invariant
masses of internal particles: it becomes lower than 40\% of the
leading-order (LO) contribution for momentum transfer squared
$Q^2>7$ GeV$^2$. It is observed that the NLO leading-twist
correction does not play an essential role in explaining the
experimental data, but the LO higher-twist contribution does.

\end{abstract}


\maketitle

\section{INTRODUCTION}

The $k_T$ factorization theorem \cite{CCH,CE,LRS,BS,LS,HS} in
perturbative QCD (PQCD) has caught a lot of attention recently
(see \cite{DXY10,BLZ10} for its recent applications). How
to derive a hard kernel beyond leading order (LO) in the $k_T$
factorization theorem is one of the important subjects. We have
proposed the prescription that partons in both QCD quark diagrams
and effective diagrams for transverse-momentum-dependent (TMD)
hadron wave functions are off mass shell by $k_T^2$ \cite{NL2}.
Since the gauge dependence cancels between the two sets of diagrams,
their difference defines a gauge-invariant hard kernel. The gauge
invariance of the $k_T$ factorization theorem has been investigated
in \cite{NL07,FMW08}, and proved in \cite{LM09}. The light-cone
singularities, appearing in the naive definition of TMD wave
functions, are regularized by rotating the Wilson lines away from
the light cone \cite{Co03,LL04,MW}. Following the above
prescription, the next-to-leading-order (NLO) correction to the pion
transition form factor associated with the process $\pi\gamma^*\to
\gamma$ has been calculated at leading twist \cite{NL07}. Both the
large double logarithms $\alpha_s\ln^2k_T$ and $\alpha_s\ln^2 x$,
$x$ being a parton momentum fraction, were identified. The former is
absorbed into the pion wave function and summed to all orders in
$\alpha_s$ by the $k_T$ resummation, and the latter is absorbed
into a jet function and summed to all orders in
$\alpha_s$ by the threshold resummation.

In this paper we shall extend the framework to the more
complicated pion electromagnetic form factor associated with the
process $\pi\gamma^*\to \pi$. Our work represents the first NLO
study of this quantity at leading twist in the $k_T$ factorization
theorem. The NLO correction to the pion form factor has been
evaluated in the collinear factorization theorem by several groups
\cite{FGO81,DR81,S82,KR85,BT87,KMR86,MNP99,Braun:1999uj,Bijnens:2002mg}.
For $\pi\gamma^*\to \pi$, not only the collinear divergences from
gluon emissions collimated to the initial- and final-state pions,
but also the soft divergences from gluon exchanges between the two
pions exist. Therefore, it is nontrivial to demonstrate that the
collinear divergences in the quark diagrams are cancelled by those
in the pion wave functions, and the soft divergences cancel among
themselves. The computation of the box and pentagon diagrams for
the pion form factor is also technically involved. The
renormalization scale $\mu$ and the factorization scale $\mu_{\rm
f}$ are introduced by higher-order corrections to the quark
diagrams and to the effective diagrams, respectively. We shall
show that the NLO correction to the pion form factor is smaller,
when both the scales are set to the invariant masses of internal
particles as postulated in \cite{LS,KLS}: it becomes lower than
40\% of the LO contribution for momentum transfer squared $Q^2>7$
GeV$^2$. The variation of the pion form factor with $\mu$ and
$\mu_{\rm f}$ gives an estimate of the theoretical uncertainty in
our analysis.

Another related subject concerns the shape of the pion distribution
amplitudes. It has been pointed out \cite{LM092} that the recent
BaBar data of the pion transition form factor $F_{\pi\gamma}(Q^2)$
\cite{BABAR} at low (high) $Q^2$ indicate an asymptotic (flat)
leading-twist, i.e., twist-2 pion distribution amplitude. However,
a pion distribution amplitude ought to be universal. These
seemingly contradictory observations have been reconciled in the
$k_T$ factorization theorem \cite{LM092}: the increase of the
measured $Q^2F_{\pi\gamma}(Q^2)$ for $Q^2> 10$ GeV$^2$ was explained
by convoluting a $k_T$-dependent hard kernel with a flat pion
distribution amplitude. The low $Q^2$ data were accommodated by
including the threshold resummation of $\alpha_s\ln^2x$, which
provides a strong suppression at the endpoints of $x$. The suppression
becomes stronger as $Q^2$ decreases. That is, the
twist-2 pion distribution amplitude multiplied by the threshold
resummation factor behaves like the asymptotic model at low $Q^2$
effectively. In this work
we shall show that the asymptotic models for both the twist-2 and
two-parton twist-3 pion distribution amplitudes also lead to results
in better agreement with the data of the pion form factor at low
$Q^2$, compared to those from the non-asymptotic models. It is found
that the NLO twist-2 correction does not play an essential role in
accounting for the data, but the contribution from the LO two-parton
twist-3 pion distribution amplitudes does, a conclusion the same as
drawn in \cite{CKO09}.

In Sec.~II we calculate the $O(\alpha_s^2)$ QCD quark diagrams, the
$O(\alpha_s)$ effective diagrams for the TMD pion wave function, and
their convolution with the $O(\alpha_s)$ hard kernel, considering
partons off-shell by $k_T^2$. Since the $k_T$ factorization theorem
is appropriate for QCD processes dominated by contributions from
small $x$ \cite{NL2}, we shall keep only terms in leading power of
$x$. Taking the difference between the two sets of diagrams, we
derive the NLO $k_T$-dependent hard kernel for the pion form factor.
Section III contains the numerical investigation, in which we obtain
the partial contributions to the pion form factor from different
twists and orders, and the relative importance of the LO and NLO
twist-2 contributions. We also examine the dependence of our results
on the renormalization and factorization scales, and on the shape of
the pion distribution amplitudes. Section IV is the conclusion.

\section{NLO CORRECTIONS}

\begin{figure}[t]
\begin{center}
\hspace{-3 cm}
\includegraphics[height=7cm]{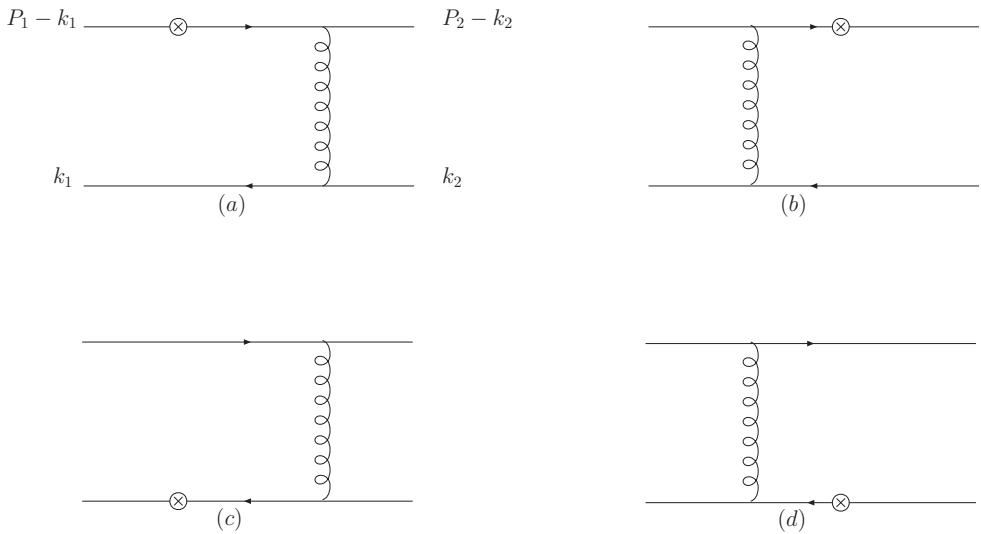}
\caption{Leading-order quark diagrams for $\pi\gamma^*\to\pi$ with
$\otimes$ representing the virtual photon vertex.} \label{leading}
\end{center}
\end{figure}

In this section we compute the $O(\alpha_s^2)$ quark diagrams in QCD
and the $O(\alpha_s)$ effective diagrams for the pion wave function
in the Feynman gauge,
and then derive the NLO hard kernel for the pion electromagnetic
form factor in the $k_T$ factorization theorem. The momentum $P_1$
($P_2$) of the initial-state (final-state) pion is chosen as $P_1
=(P_1^+,0,{\bf 0}_T)$ ($P_2 = (0,P_2^-,{\bf 0}_T)$). The anti-quark
$\bar q$ carries the momentum $k_1=(x_1P_1^+,0,{\bf k}_{1T})$ in the
initial state and $k_2=(0,x_2P_2^-,{\bf k}_{2T})$ in the final
state, as indicated by the LO quark diagrams in Fig.~\ref{leading}.
The parton virtuality $k_T^2$ regularizes the collinear and soft
divergences into the logarithm $\ln k_T^2$. We shall focus only on
Fig.~\ref{leading}(a), since the NLO correction to
Fig.~\ref{leading}(b) can be obtained from that to
Fig.~\ref{leading}(a) by exchanging the kinematic variables of the
two pions. The NLO corrections to the other two LO quark diagrams with
the virtual photon attaching to the anti-quark line are then obtained
via the variable exchanges between the quark and the anti-quark.
According to the hierarchy $Q^2\gg xQ^2\gg k_{T}^2$ in
the small-$x$ region, we keep only those terms that do not vanish in
the $x\to 0$ and $k_T\to 0$ limit.

Figure~\ref{leading}(a) leads to the amplitude
\begin{eqnarray}
H^{(0)}(x_1,k_{1T},x_2,k_{2T},Q^2)&=&-g^2C_F\frac{N_c}{(\sqrt{2N_c})^2}
\frac{Tr[\gamma_\nu\gamma_5\not\! P_2\gamma^\nu(\not\! P_2-\!\not
k_1)\gamma_\mu\not\!
P_1\gamma_5]}{(P_2-k_1)^2(k_1-k_2)^2},\nonumber\\
%&=&g^2C_F\frac{x_1Tr(\not\! P_2\not\! P_1\gamma_\mu\not\! P_1)}{(x_1
%Q^2+k_{1T}^2)(x_1 x_2Q^2+|{\bf k}_{1T}-{\bf k}_{2T}|^2)},\nonumber\\
&=&g^2C_F\frac{Tr(\not\! P_2\not\! P_1\gamma_\mu\not\! P_1)}{
Q^2(x_1 x_2Q^2+|{\bf k}_{1T}-{\bf k}_{2T}|^2)}, \label{fd0}
\end{eqnarray}
where $N_c$ is the number of colors, $C_F$ is a color factor,
$\not\!P_1\gamma_5/\sqrt{2N_c}$ and $\gamma_5\not\!P_2/\sqrt{2N_c}$
are the twist-2 spin structures of the initial- and final-state
pions, respectively, and $Q^2\equiv 2P_1\cdot P_2$ is the momentum
transfer squared from the virtual photon. To reach the last line in
Eq.~(\ref{fd0}), we have applied the hierarchy $x_1Q^2\gg k_{1T}^2$
to the internal quark propagator. The denominator $x_1x_2Q^2+|{\bf
k}_{1T}-{\bf k}_{2T}|^2$ comes from the virtuality of the LO hard
gluon. The $|{\bf k}_{1T}-{\bf k}_{2T}|^2$ term may not be
negligible compared to $x_1 x_2Q^2$ in the small-$x$ region, so it
is retained.

The NLO hard kernel $H^{(1)}$ is defined, in the $k_T$ factorization
theorem, by \cite{NL2}
\begin{eqnarray}
H^{(1)}(x_1,k_{1T},x_2,k_{2T},Q^2)&=&G^{(1)}(x_1,k_{1T},x_2,k_{2T},Q^2)\nonumber\\
& &-\int dx'_1d^2k'_{1T}\Phi^{(1)}(x_1,k_{1T};x'_1,k'_{1T})
H^{(0)}(x'_1,k'_{1T},x_2,k_{2T},Q^2)\nonumber\\
& &-\int dx'_2 d^2k'_{2T}H^{(0)}(x_1,k_{1T},x'_2,k'_{2T},Q^2)
\Phi^{(1)}(x_2,k_{2T};x'_2,k'_{2T}), \label{pa1}
\end{eqnarray}
where $G^{(1)}$ denotes the NLO quark diagrams associated with
Fig.~\ref{leading}(a), and $\Phi^{(1)}$ collects the $O(\alpha_s)$
effective diagrams for the twist-2 quark-level wave function
\cite{NL2,L1}
\begin{eqnarray}
\Phi(x_1,k_{1T};x'_1,k'_{1T})=\int\frac{dy^-}{2\pi
}\frac{d^2y_T}{(2\pi)^2}e^{-ix'_1P_1^+ y^-+i{\bf k}'_{1T}\cdot {\bf
y}_T}\langle 0|{\bar q}(y) W_y(n)^{\dag}I_{n;y,0}W_0(n) \not
n_-\gamma_5 q(0)|q(P_1-k_1)\bar q(k_1)\rangle.\label{de1}
\end{eqnarray}
In the above expression $y=(0,y^-,{\bf y}_T)$ is the coordinate of
the anti-quark field $\bar q$, $n_-=(0,1,{\bf 0}_T)$ is a null
vector along $P_2$, $|q(P_1- k_1)\bar q(k_1)\rangle$ is the leading
Fock state of the pion, and the Wilson line $W_y(n)$ with
$n^2\not=0$ is written as
\begin{eqnarray}
\label{eq:WL.def} W_y(n) = P \exp\left[-ig \int_0^\infty d\lambda
n\cdot A(y+\lambda n)\right].
\end{eqnarray}
The two Wilson lines $W_y(n)$ and $W_0(n)$ are connected by a
vertical link $I_{n;y,0}$ at infinity \cite{BJY,CS08}.
Equation~(\ref{de1}) will generate additional light-cone
singularities from the region with a loop momentum collinear to
$n_-$, as the Wilson line direction approaches the light cone, i.e.,
as $n\to n_-$ \cite{Co03}. That is, $n^2$ serves as an infrared
regulator for the light-cone singularities.

\subsection{NLO Quark Diagrams}

\begin{figure}[t]
\begin{center}
\hspace{-2 cm}
\includegraphics[height=7cm]{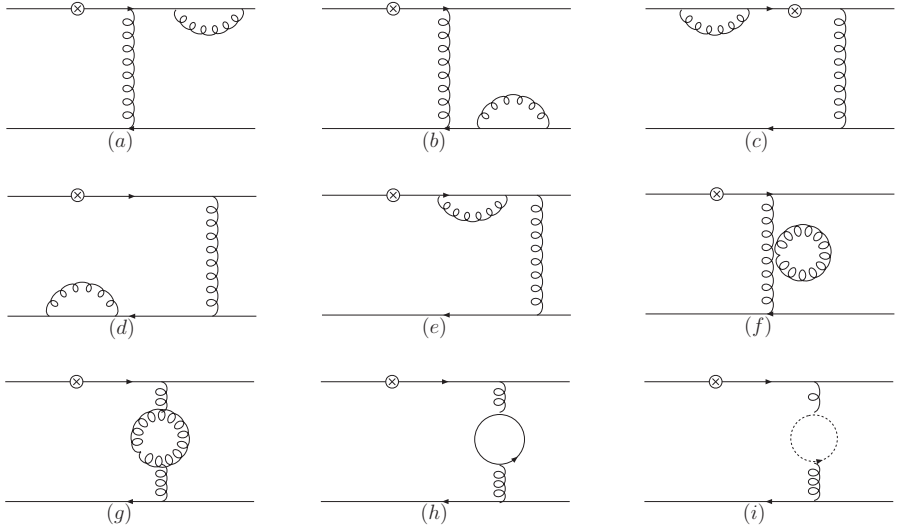}
\caption{Self-energy diagrams.} \label{selfd}
\end{center}
\end{figure}

\begin{figure}[t]
\begin{center}
\hspace{-2 cm}
\includegraphics[height=7cm]{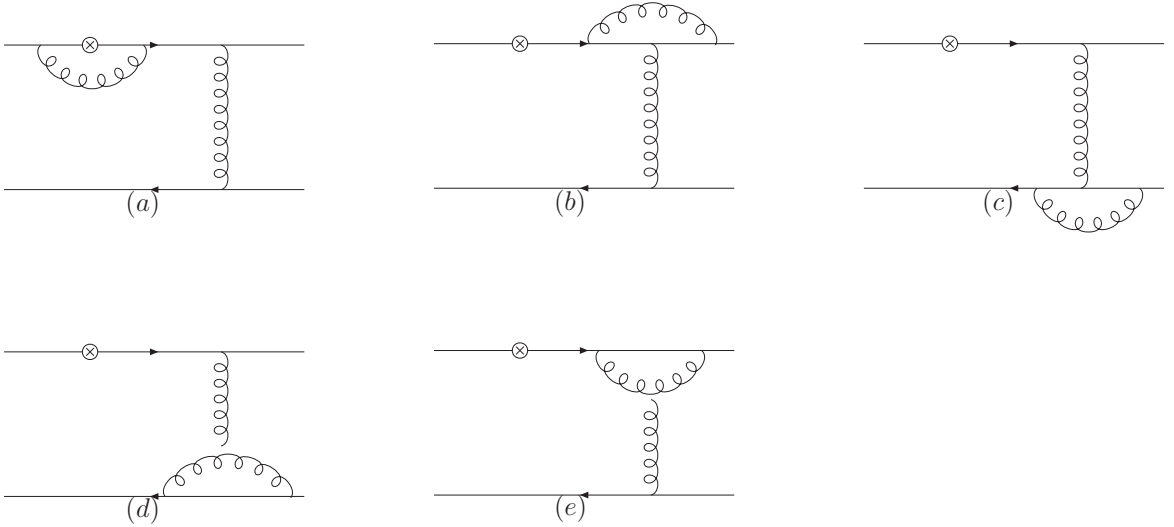}
\caption{Vertex-correction diagrams.} \label{three}
\end{center}
\end{figure}

\begin{figure}[t]
\begin{center}
\hspace{-3 cm}
\includegraphics[height=7cm]{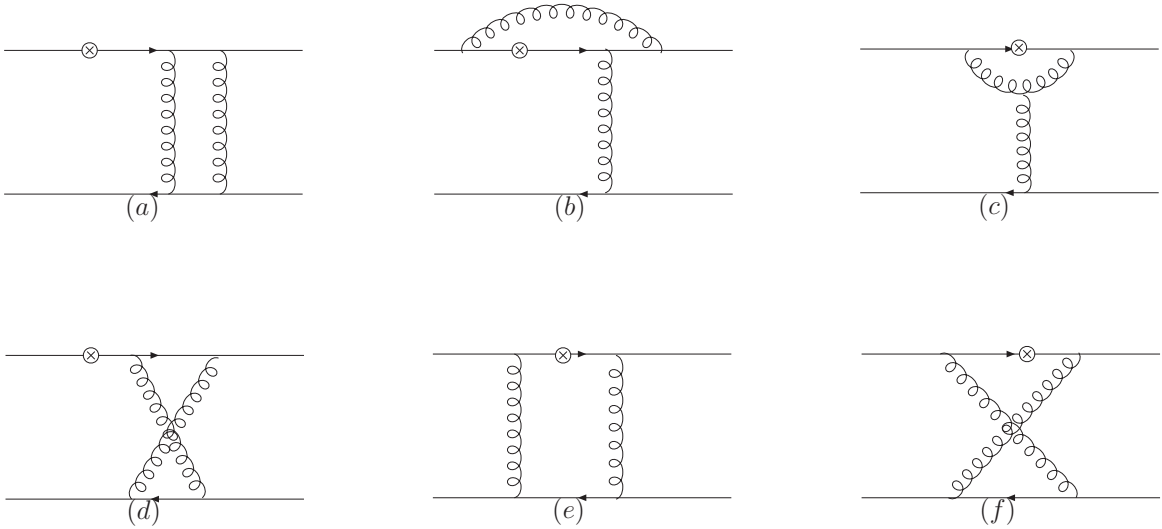}
\caption{Box and pentagon diagrams.} \label{four}
\end{center}
\end{figure}

We first calculate the NLO corrections to Fig.~\ref{leading}(a) in the
$k_T$ factorization theorem, which come from Figs.~\ref{selfd},
\ref{three}, and \ref{four} for the self-energy corrections, the
vertex corrections, and the box and pentagon diagrams, respectively.
The ultraviolet poles are identified in the dimensional reduction
\cite{WS79} in order to avoid the ambiguity from handling the matrix
$\gamma_5$. To simplify the expressions, we define the following
dimensionless ratios
\begin{eqnarray}
& &\delta_{1} =\frac{k_{1T}^2}{Q^2},\;\;\;\; \delta_{2}
=\frac{k_{2T}^2}{Q^2},\nonumber\\
& &\delta_{12} =\frac{x_1 x_2 Q^2 + {|{\bf k}_{1T}-{\bf
k}_{2T}|}^2}{Q^2}.
\end{eqnarray}
Terms suppressed by powers of $x$ or $\delta$ will be dropped in the
NLO corrections.

The self-energy corrections to the four external quarks in
Figs.~\ref{selfd}(a)-\ref{selfd}(d) give
\begin{eqnarray}
G^{(1)}_{2a,2b} &=&-\frac{\alpha_sC_F}{8\pi}\left(\frac{1}{\epsilon}
+\ln\frac{4\pi\mu^2}{\delta_1 Q^2e^{\gamma_E}}+2\right)
H^{(0)},\label{pga1}\\
G^{(1)}_{2c,2d}&=& -\frac{\alpha_sC_F}{8\pi}\left(\frac{1}{\epsilon}
+\ln\frac{4\pi\mu^2}{\delta_2Q^2 e^{\gamma_E}}+2\right) H^{(0)},
\label{pgb1}
\end{eqnarray}
where $1/\epsilon$ represents the ultraviolet pole, $\mu$ is the
renormalization scale, and $\gamma_E$ is the Euler constant. The
collinear divergences in Figs.~\ref{selfd}(a)-\ref{selfd}(d) are
regularized into the infrared logarithms $\ln \delta$. The
self-energy correction to the internal quark in Fig.~\ref{selfd}(e)
leads to
\begin{eqnarray}
G^{(1)}_{2e}=-\frac{\alpha_sC_F}{4\pi}
\left(\frac{1}{\epsilon}+\ln\frac{4\pi\mu^2 }{x_1Q^2
e^{\gamma_E}}+2\right) H^{(0)}.\label{pgc1}
\end{eqnarray}
The ultraviolet poles in Eqs.~(\ref{pga1}) and (\ref{pgb1}) are half
of that in Eq.~(\ref{pgc1}), because an additional factor $1/2$ is
associated with an external particle. The internal quark is
off-shell by the invariant mass squared $x_1Q^2$, which replaces the
arguments $k_T^2$ of the infrared logarithms in Eqs.~(\ref{pga1})
and (\ref{pgb1}). The self-energy correction to the hard gluon,
summing the quark-, gluon- and ghost-loop contributions in
Figs.~\ref{selfd}(f)-\ref{selfd}(i), is written as
\begin{eqnarray}
G^{(1)}_{2f+2g+2h+2i}=\frac{\alpha_s}{4\pi}
\left(\frac{5}{3}N_c-\frac{2}{3}N_f\right)\left(\frac{1}{\epsilon}+\ln\frac{4\pi\mu^2
}{\delta_{12}Q^2e^{\gamma_E}}\right) H^{(0)},\label{pg1}
\end{eqnarray}
with $N_f$ being the number of quark flavors. In this case the
logarithm depends on the invariant mass squared $\delta_{12}Q^2$ of
the hard gluon.

The results from the five diagrams
Figs.~\ref{three}(a)-\ref{three}(e) for the vertex corrections are
summarized as
\begin{eqnarray}
G_{3a}^{(1)} &=& \frac{\alpha_sC_F}{4\pi} \bigg
(\frac{1}{\epsilon}+\ln\frac{4\pi\mu^2}{Q^2e^{\gamma_E}} +2
\bigg)H^{(0)},
\\ G_{3b}^{(1)}&=&
-\frac{\alpha_s}{ 8 \pi N_c} \bigg
(\frac{1}{\epsilon}+\ln\frac{4\pi\mu^2}{x_1 Q^2e^{\gamma_E}}
+2\ln\frac{x_1}{\delta_2} +\frac{7}{2} \bigg )H^{(0)},
\\
G_{3c}^{(1)} &=&-\frac{\alpha_s}{ 8 \pi N_c} \bigg
(\frac{1}{\epsilon}+\ln\frac{4\pi\mu^2}{\delta_{12}
Q^2e^{\gamma_E}}
 - 2 \ln{ \delta_{12} \over  \delta_1}  \ln { \delta_{12} \over
\delta_2 }  + 2 \ln{ \delta_{12}^2 \over   \delta_1 \delta_2} -{2
\pi^2 \over 3 } +\frac{11}{2}   \bigg)H^{(0)},
\\
G_{3d}^{(1)}&=& \frac{\alpha_s N_c}{8 \pi} \bigg
(\frac{3}{\epsilon}+ 3 \ln\frac{4\pi\mu^2}{\delta_{12}
Q^2e^{\gamma_E}} + 2 \ln { \delta_{12}^2 \over \delta_1 \delta_2
}+\frac{23}{2}  \bigg )H^{(0)},
 \\
G_{3e}^{(1)}&=&\frac{\alpha_s N_c }{8\pi}\bigg
[\frac{3}{\epsilon}+3\ln\frac{4\pi\mu^2}{x_1 Q^2e^{\gamma_E}}
+2\ln\frac{x_1}{\delta_2}\left(1-
\ln\frac{x_1}{\delta_{12}}\right)+\ln\frac{x_1}{\delta_{12}}-\frac{2}{3}\pi^2
+\frac{7}{2}\bigg ]H^{(0)}.\label{3e}
\end{eqnarray}
The ultraviolet poles in the self-energy and vertex corrections are
summed into
\begin{eqnarray}
%& &\frac{\alpha_s}{4\pi}\left(-C_F-C_F-C_F
%+\frac{5}{3}N_c-\frac{2}{3}N_f+C_F
%-\frac{1}{2N_c}-\frac{1}{2N_c}+\frac{3}{2}N_c+\frac{3}{2}N_c\right)
%\frac{1}{\epsilon}
%\nonumber\\&=&
\frac{\alpha_s}{4\pi}\left(11-\frac{2}{3}N_f\right)\frac{1}{\epsilon},
\end{eqnarray}
for $N_c=3$, which is the same as obtained in \cite{MNP99}. This pole term
determines the renormalization-group (RG) evolution of the coupling
constant $\alpha_s$.

The loop correction $G_{3a}^{(1)}$ to the photon vertex does not
contain an infrared logarithm, because of the sequence of the
$\gamma$-matrices in the fermion flow. The correction $G_{3b}^{(1)}$
to the upper gluon vertex contains only the infrared logarithm
associated with the final-state pion. The other end of the radiative
gluon attaches to the internal quark line, so $G_{3b}^{(1)}$ depends
on $x_1Q^2$, and does not generate a double logarithm. The
correction to the lower gluon vertex in Fig.~\ref{three}(c) depends
on the momentum transfer squared $\delta_{12}Q^2$ from the LO hard
gluon, and on the two infrared regulators $\delta_1$ and $\delta_2$
from the incoming and outgoing partons. The expression of
$G^{(1)}_{3c}$ is symmetric under the exchange of $\delta_1$ and
$\delta_2$ as expected. The
$\ln(\delta_{12}/\delta_1)\ln(\delta_{12}/\delta_2)$ term is not a
large double logarithm in the small-$x$ region, where
$\ln\delta_1\ln\delta_2$ arises from the soft region of the
radiative gluon. It is obvious from the convolutions of $\Phi^{(1)}$
and $H^{(0)}$ in Eq.~(\ref{pa1}) that this soft logarithm can not be
cancelled by the infrared logarithms in the effective diagrams:
there is no chance for $\ln\delta_1$ associated with the
initial-state pion and $\ln\delta_2$ associated with the final-state
pion to appear in a product. Therefore, cancellation must occur
within this type of soft logarithms as demonstrated below.

The triple-gluon vertex correction $G_{3d}^{(1)}$ also depends on
the scale $\delta_{12}Q^2$, and contains the same infrared logarithm
$\ln(\delta_{12}^2/\delta_1 \delta_2)$ as in $G_{3c}^{(1)}$ but with
different color factors. Their sum is proportional to $C_F$:
$-1/N_c+N_c=2C_F$, implying the factorization of this infrared
logarithm in color flow. It is then possible that it will be
cancelled by the corresponding logarithm in the effective diagrams,
whose color factor is also $C_F$. This combination of quark diagrams
for achieving factorization also takes place elsewhere. Another
triple-gluon vertex correction in Fig.~\ref{three}(e) involves the
invariant masses squared of the virtual quark and of the hard
gluon as shown in Eq.~(\ref{3e}). At small $x$, the quark $q$ in
Fig.~\ref{three}(e) is energetic, leading to the collinear
logarithmic enhancement $\ln(x_1/\delta_2)$, and the hard gluon has
a small invariant mass, leading to the soft enhancement
$\ln(x_1/\delta_{12})$. Their overlap then gives rise to the
important double logarithm $\ln(x_1/\delta_2) \ln(x_1/\delta_{12})$
in $G^{(1)}_{3e}$.
%In the region with $x_2\sim O(1)$, the hard gluon
%becomes off-shell by $O(x_1Q^2)$, the soft enhancement disappears
%like $\ln(x_1/\delta_{12})\sim O(1)$, and the double logarithm
%reduces to a single logarithm. $G^{(1)}_{3e}$ clearly exhibits the
%transition of the double logarithm at small $x_2$ to the single
%logarithm at large $x_2$.
This double logarithm can be
reexpressed as
\begin{eqnarray}
2\ln\frac{x_1}{\delta_2}\ln\frac{x_1}{\delta_{12}}=
\ln^2\frac{x_1}{\delta_2} +\ln^2\frac{\delta_{12}}{x_1}
-\ln^2\frac{\delta_{12}}{\delta_2},\label{dou}
\end{eqnarray}
where the first term, known as the Sudakov logarithm \cite{BS,CS},
is absorbed into the final-state pion wave function. The Sudakov
effect from resumming this double logarithm suppresses the
contribution from the small $k_{2T}$ region, i.e., the region with a
large impact parameter \cite{LS}. The second term
$\ln^2(\delta_{12}/x_1)\approx\ln^2 x_2$ also exists in the
collinear factorization theorem with $k_T$ being integrated out
\cite{KPY,MW0607,ASY}. This threshold logarithm is significant at
small $x_2$, where the hard gluon approaches mass shell \cite{UL}.
The third term in Eq.~(\ref{dou}) is less important in the small-$x$
region.

There are four box diagrams and two pentagon diagrams as displayed
in Fig.~\ref{four}. Note that Figs.~\ref{four}(c), \ref{four}(e)
and \ref{four}(f) also contain the corrections to the LO quark
diagram Fig.~\ref{leading}(b). As explained in \cite{NL07}, the
two-particle reducible diagrams, Figs.~\ref{four}(a) and
\ref{four}(e), give power-suppressed contributions at small $x$.
The reason is simply that a collinear or ultraviolet loop momentum
increases the virtuality of the LO hard gluons. A soft loop
momentum does not change the power-law behavior of the hard
kernel, which is, however, not a leading region due to the soft
cancellation mentioned before. The expressions for the diagrams
other than Figs.~\ref{four}(a) and \ref{four}(e) are collected
below:
\begin{eqnarray}
G^{(1)}_{4b}&=&\frac{\alpha_s }{4\pi N_c} \bigg [\ln\delta_1 (1+
\ln \delta_2) + 2 \ln x_1 +
\frac{\pi^2}{3} - 1 \bigg ]H^{(0)}, \\
G^{(1)}_{4c}&=&-\frac{\alpha_s N_c }{4 \pi}  \bigg [ \ln \delta_1
(\ln x_1 +1) - \ln x_2 \left(\frac{3}{2} \ln x_1
+\frac{7}{4}\right) \bigg ]H^{(0)} -\frac{\alpha_s N_c }{4\pi}
\bigg [ x_1 \leftrightarrow x_2, \delta_1
\leftrightarrow \delta_2  \bigg ]\bar{H}^{(0)},\\
G^{(1)}_{4d}&=&-\frac{\alpha_s  }{4 \pi N_c}  \bigg ( \ln {
\delta_{12} \over \delta_{1}} \ln {x_1 \over
\delta_{2}}  +  {\pi^2 \over 6}  \bigg )H^{(0)}, \\
G_{4f}^{(1)}&=& - \frac{\alpha_s }{4\pi N_c}  \bigg ( \ln {
\delta_{12}\over x_1\delta_1} \ln \frac{\delta_{12}}{\delta_2}-\ln
4-{\pi^2 \over 4} -{1 \over 2} \bigg )H^{(0)}- \frac{\alpha_s }{4\pi
N_c} \bigg(x_1 \leftrightarrow x_2, \delta_1 \leftrightarrow
\delta_2 \bigg )\bar{H}^{(0)},
\end{eqnarray}
with $\bar{H}^{(0)}$ being the LO hard kernel from
Fig.~\ref{leading}(b)
\begin{eqnarray}
\bar{H}^{(0)}(x_1,x_2,k_{1T},k_{2T},Q^2)
%&=&g^2C_F\frac{ x_2Tr[\!\not
%P_1\not P_2\gamma_\mu\!\not P_2]}{(x_2Q^2+k_{2T}^2)(x_1x_2Q^2+|{\bf
%k}_{1T}-{\bf k}_{2T}|^2)},\nonumber\\
&=&g^2C_F\frac{Tr(\!\not P_1\not P_2\gamma_\mu\!\not P_2)}{
Q^2(x_1x_2Q^2+|{\bf k}_{1T}-{\bf k}_{2T}|^2)}.\label{hbar}
\end{eqnarray}

The above expressions are free of ultraviolet divergences. The
correction $G^{(1)}_{4c}$ involves only $\ln\delta_1$, if focusing
on the correction to Fig.~\ref{leading}(a), and the important double
logarithm $\ln\delta_1\ln x_1$: $\ln\delta_1$ denotes the collinear
enhancement, and $\ln x_1$ denotes the soft enhancement in the small
$x_1$ region. Similarly, the above double logarithm can be
reexpressed as
\begin{eqnarray}
2\ln\delta_1\ln x_1=\ln^2\delta_1 +\ln^2
x_1-\ln^2\frac{x_1}{\delta_1},\label{double}
\end{eqnarray}
for the separation the Sudakov logarithm and the threshold
logarithm. Figures.~\ref{four}(b), \ref{four}(d), and \ref{four}(f),
with the soft radiative gluons attaching to the incoming and
outgoing partons, generate the soft logarithm
$\ln\delta_1\ln\delta_2$. These soft logarithms cancel among
Figs.~\ref{three}(c), \ref{four}(b), \ref{four}(d), and
\ref{four}(f), for the reason that soft gluons do not interact with
color-singlet objects like pions.

The sum over all the NLO quark diagrams associated with
Fig.~\ref{leading}(a) gives
\begin{eqnarray}
G^{(1)}
&=&{ \alpha_s C_F  \over 4 \pi }\bigg[ {21 \over 4}\bigg({1 \over
\epsilon} + \ln {4 \pi \mu^2 \over Q^2 e^{\gamma_E}} \bigg) -3 \ln
(\delta_1 \delta_2)  - {9 \over 4}\ln^2 x_1  + {27 \over 8} \ln
x_1 \ln x_2    + {9 \over 4} \ln x_1 \ln \delta_{12} \nonumber
\\
&& + \ln x_1 \bigg( 2 \ln { \delta_2 } -2 \ln \delta_1 +{3 \over
8}\bigg ) +{63 \over 16} \ln x_2 - \ln \delta_{12} \bigg( 2
\ln\delta_2 + {9 \over 8} \bigg)+{1 \over 2} \ln 2 -{7 \over 12}
\pi^2 + {69 \over 4} \bigg ]H^{(0)} , \label{pgt}
\end{eqnarray}
for $N_f=6$. The NLO correction to Fig.~\ref{leading}(b)
can be obtained from the above expression by performing the
exchanges $x_1\leftrightarrow x_2$ and
$\delta_1\leftrightarrow\delta_2$.

\subsection{NLO Effective Diagrams}

We compute the convolutions of the NLO wave functions $\Phi^{(1)}$
with the LO hard kernel $H^{(0)}$ over the integration variables
$x'$ and $k'_T$,
\begin{eqnarray}
\Phi^{(1)}_{i}\otimes H^{(0)}&\equiv& \int dx'_1 d^2k'_{1T}
\Phi^{(1)}_i(x_1,k_{1T};x'_1,k'_{1T})H^{(0)}(x'_1,k'_{1T},x_2,k_{2T},Q^2),
\nonumber\\
H^{(0)}\otimes \Phi^{(1)}_{i}&\equiv& \int dx'_2 d^2k'_{2T}
H^{(0)}(x_1,k_{1T},x'_2,k'_{2T},Q^2)\Phi^{(1)}_i(x_2,k_{2T};x'_2,k'_{2T}),
\end{eqnarray}
with a choice of $n^2\not=0$ for the
Wilson line direction to regularize the light-cone singularities.
The NLO wave functions then depend on $n^2$ through the scale
$\zeta_1^2\equiv 4(n \cdot P_1)^2/|n^2|$ or $\zeta_2^2\equiv 4(n
\cdot P_2)^2/|n^2|$, which is regarded as a factorization-scheme
dependence. This dependence, entering a hard kernel when taking the
difference between the quark diagrams and the effective diagrams,
can be minimized by adhering to a fixed $n^2$. Note that the soft
subtraction factor introduced in \cite{FMW08} is not necessary here,
which is needed for a choice of $n^2=0$.

\begin{figure}[t]
\begin{center}
\hspace{-2 cm}
\includegraphics[height=6 cm]{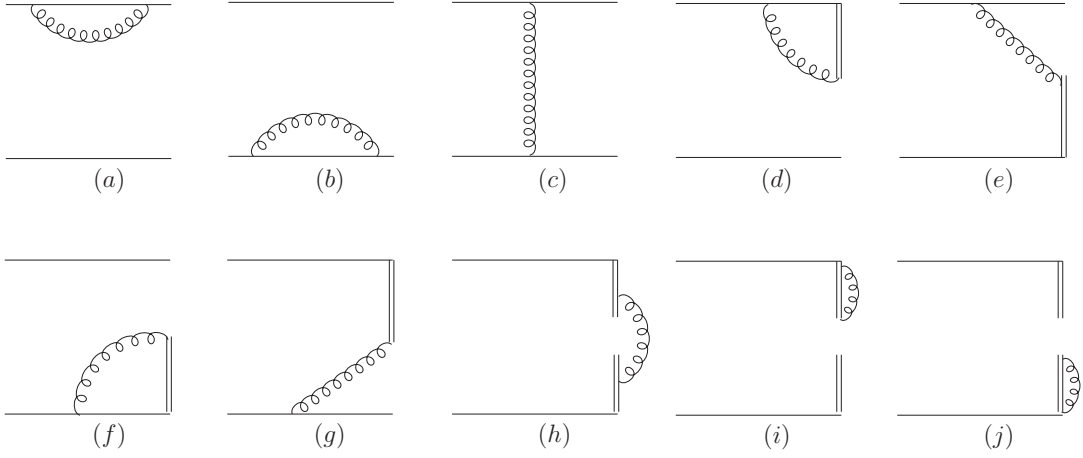}
\caption{Effective diagrams.} \label{ktlc2}
\end{center}
\end{figure}

The self-energy corrections from Figs.~\ref{ktlc2}(a) and
\ref{ktlc2}(b) are written as
\begin{eqnarray}
\Phi^{(1)}_{a}\otimes H^{(0)}=\Phi^{(1)}_{b}\otimes H^{(0)}
&=&-\frac{\alpha_s C_F}{8\pi}\left(\frac{1}{\epsilon}
+\ln\frac{4\pi\mu_{\rm
f}^2}{\delta_1Q^2e^{\gamma_E}}+2\right)H^{(0)},
\label{pwa}\\
H^{(0)}\otimes\Phi^{(1)}_{a}=
H^{(0)}\otimes\Phi^{(1)}_{b}&=&-\frac{\alpha_s
C_F}{8\pi}\left(\frac{1}{\epsilon} +\ln\frac{4\pi\mu_{\rm
f}^2}{\delta_2Q^2e^{\gamma_E}}+2\right)H^{(0)},\label{pwb}
\end{eqnarray}
whose expressions are similar to those from the quark diagrams but
with the factorization scale $\mu_{\rm f}$. The contribution from
the box diagram in Fig.~\ref{ktlc2}(c) is power-suppressed in the
small $x$ region:
\begin{eqnarray}
\Phi^{(1)}_{c}\otimes H^{(0)}=H^{(0)}\otimes\Phi^{(1)}_{c}=0 .
\end{eqnarray}

The sign of the plus component $n^+$ of the vector $n$ is arbitrary,
which could be positive or negative ($n^-$ has a positive sign, the
same as of $P_2^-$). Choosing $n^+<0$, i.e., $n^2<0$ as in
\cite{LS,LY1,Li96}, Fig.~\ref{ktlc2}(d) leads, in the small $x$
region, to
\begin{eqnarray}
\Phi^{(1)}_{d}\otimes
H^{(0)}=\frac{\alpha_sC_F}{4\pi}\left(\frac{1}{\epsilon}
+\ln\frac{4\pi\mu_{\rm f}^2}{k_{1T}^2e^{\gamma_E}}
-\ln^2\frac{\zeta_1^2}{k_{1T}^2}+\ln\frac{\zeta_1^2}{k_{1T}^2}+2-\frac{\pi^2}{3}
\right) H^{(0)},\nonumber\\
H^{(0)}\otimes\Phi^{(1)}_{d}=\frac{\alpha_sC_F}{4\pi}\left(\frac{1}{\epsilon}
+\ln\frac{4\pi\mu_{\rm f}^2}{k_{2T}^2e^{\gamma_E}}
-\ln^2\frac{\zeta_2^2}{k_{2T}^2}+\ln\frac{\zeta_2^2}{k_{2T}^2}+2-\frac{\pi^2}{3}
\right) H^{(0)},\label{pwd}
\end{eqnarray}
which reproduces the Sudakov logarithm in the form of $\ln^2
(\zeta^2/k_T^2)$. As computing the convolution of $\Phi_e^{(1)}$
with $H^{(0)}$, the momentum fraction appearing in the hard kernel
should be restricted between 0 and 1. The expression for
Fig.~\ref{ktlc2}(e) is given, in the small $x$ region, by
\begin{eqnarray}
\Phi^{(1)}_{e}\otimes H^{(0)}&=& {\alpha_sC_F  \over 4\pi } \bigg [
\ln^2 \bigg({x_1 \zeta_1^2\over k_{1T}^2} \bigg ) + { 2 \pi^2 \over
3}
\bigg ] H^{(0)},\nonumber\\
H^{(0)}\otimes \Phi^{(1)}_{e}&=& \frac{\alpha_sC_F}{4\pi} \bigg [
\ln^2 \bigg({\delta_{12} \zeta_2^2\over x_1k_{2T}^2} \bigg ) + { 2
\pi^2 \over 3} \bigg ] H^{(0)}, \label{pwe}
\end{eqnarray}
where terms vanishing with $k_{T}^2\to 0$ have been dropped. It is
observed that Fig.~\ref{ktlc2}(e) also generates a double logarithm,
whose importance is attenuated by the small $x$.

The result from Fig.~\ref{ktlc2}(f) is similar to that from
Fig.~\ref{ktlc2}(d), but with the replacement of $P-k$ by $k$, i.e.,
$\zeta$ by $x\zeta$. Keeping terms which do not vanish with
$k_T^2\to 0$, we have
\begin{eqnarray}
\Phi^{(1)}_{f}\otimes
H^{(0)}&=&\frac{\alpha_sC_F}{4\pi}\left(\frac{1}{\epsilon}
+\ln\frac{4\pi\mu_{\rm f}^2}{k_{1T}^2e^{\gamma_E}}
-\ln^2\frac{x_1^2\zeta_1^2}{k_{1T}^2}+\ln\frac{x_1^2\zeta_1^2}{k_{1T}^2}+2-\frac{\pi^2}{3}
\right)H^{(0)},\nonumber\\
H^{(0)}\otimes\Phi^{(1)}_{f}&=&\frac{\alpha_sC_F}{4\pi}\left(\frac{1}{\epsilon}
+\ln\frac{4\pi\mu_{\rm f}^2}{k_{2T}^2e^{\gamma_E}}
-\ln^2\frac{x_2^2\zeta_2^2}{k_{2T}^2}+\ln\frac{x_2^2\zeta_2^2}{k_{2T}^2}+2-\frac{\pi^2}{3}
\right)H^{(0)},\label{pwf}
\end{eqnarray}
where the double logarithm is further attenuated by $x^2$. It should
disappear, after combined with the contribution from
Fig.~\ref{ktlc2}(g), since such a double logarithm is absent in the
NLO quark diagrams. The same variable transformation relating
$\Phi^{(1)}_{f}$ to $\Phi^{(1)}_{d}$ is not applicable to
$\Phi^{(1)}_{g}$, for the latter involves the nontrivial convolution
with $H^{(0)}$. Retaining terms which are finite as $k_T\to 0$,
Fig.~\ref{ktlc2}(g) leads to
\begin{eqnarray}
\Phi^{(1)}_{g} \otimes H^{(0)}&=&  \frac{\alpha_s C_F  }{4 \pi}
\bigg (\ln^2 {x_1^2\zeta_1^2
\over k_{1T}^2 } -  {\pi^2 \over  3}  \bigg )H^{(0)} ,\nonumber\\
H^{(0)}\otimes\Phi^{(1)}_{g}&=&\frac{\alpha_sC_F}{4\pi}
\bigg(\ln^2\frac{x_2^2\zeta_2^2}{k_{2T}^2}-  {\pi^2 \over
3}\bigg)H^{(0)}. \label{pwg}
\end{eqnarray}
It is easy to see the cancellation of the double logarithms between
Eqs.~(\ref{pwf}) and (\ref{pwg}).

At last, we include the self-energy corrections to the Wilson
lines, namely, Fig.~\ref{ktlc2}(h-j):
\begin{eqnarray}
\Phi^{(1)}_{h} \otimes H^{(0)}&=&  \frac{\alpha_s C_F  }{2 \pi}
\bigg(\frac{1}{\epsilon} +\ln\frac{4\pi\mu_{\rm f}^2}{\delta_{12}Q^2
e^{\gamma_E}}\bigg )H^{(0)} ,\nonumber\\
H^{(0)}\otimes\Phi^{(1)}_{h}&=&\frac{\alpha_sC_F}{2\pi}
\bigg(\frac{1}{\epsilon} +\ln\frac{4\pi\mu_{\rm
f}^2}{\delta_{12}Q^2e^{\gamma_E}}\bigg )H^{(0)}. \label{pwh}
\end{eqnarray}
Summing all the above $O(\alpha_s)$ contributions, we obtain
\begin{eqnarray}
\Phi^{(1)}\otimes H^{(0)}&=&\sum_{i=a}^{h}\Phi^{(1)}_i\otimes
H^{(0)}\nonumber\\
&=&\frac{\alpha_sC_F}{4\pi}\left[\frac{3}{\epsilon}
+3\ln\frac{4\pi\mu_{\rm f}^2}{\zeta_1^2 e^{\gamma_E}} +(2 \ln
x_1+3)\ln\frac{\zeta_1^2}{\delta_1Q^2}+2\ln\frac{\zeta_1^2}{\delta_{12}Q^2}
+\ln x_1 (\ln x_1 +2)+2-\frac{\pi^2}{3} \right]H^{(0)},\nonumber\\
H^{(0)}\otimes \Phi^{(1)}&=&\sum_{i=a}^{h}
H^{(0)}\otimes\Phi^{(1)}_i\nonumber\\
&=&\frac{\alpha_sC_F}{4\pi}\left[\frac{3}{\epsilon}
+3\ln\frac{4\pi\mu_{\rm f}^2}{\zeta_2^2e^{\gamma_E}} +(2 \ln {
\delta_{12} \over x_1}+3)\ln\frac{\zeta_2^2}{\delta_2Q^2}
+2\ln\frac{\zeta_2^2}{\delta_{12}Q^2}+\ln^2  { \delta_{12} \over
x_1} +2 \ln x_2 +2-\frac{\pi^2}{3} \right]H^{(0)}.\nonumber\\
& &\label{ppt}
\end{eqnarray}
The resultant anomalous dimension of the pion wave
function is the same as derived in the axial gauge (for a recent
reference, see \cite{CS08}).

\subsection{NLO Hard Kernel}

We derive the NLO hard kernel for the pion electromagnetic form
factor in the $k_T$ factorization theorem by taking the difference
between the quark and effective diagrams. Note that $\alpha_s$
appearing in Eqs.~(\ref{pgt}) and (\ref{ppt}) denotes the bare
coupling constant, which can be rewritten as
\begin{eqnarray}
\alpha_s=\alpha_s(\mu_{\rm f})+\delta Z(\mu_{\rm
f})\alpha_s(\mu_{\rm f}),\label{dz}
\end{eqnarray}
with the counterterm $\delta Z$ being defined in the modified
minimal subtraction scheme. We insert Eq.~(\ref{dz}) into the
expressions of the LO and NLO quark diagrams, and of the NLO
effective diagrams. The LO hard kernel $H^{(0)}$ multiplied by
$\delta Z$ then regularizes the ultraviolet pole in Eq.~(\ref{pgt}).
The ultraviolet pole in Eq.~(\ref{ppt}) is regularized by the
counterterm of the quark field and by an additive counterterm in the
modified minimal subtraction scheme. For the infrared logarithms,
there exists a cancellation between the two-particle reducible quark
diagrams and the two-particle reducible effective diagrams. That is,
$G_{2a,2b}^{(1)}-H^{(0)}\otimes\Phi_{a,b}^{(1)}$ and
$G_{2c,2d}^{(1)}-\Phi_{a,b}^{(1)}\otimes H^{(0)}$ are infrared
finite. The self-energy diagrams for the internal quarks and gluons
are free of infrared divergences. The collinear logarithm
represented by $\ln\delta_1$ ($\ln\delta_2$) cancels between the
two-particle irreducible diagrams and the convolution
$[\Phi_d^{(1)}+\Phi_e^{(1)}+\Phi_f^{(1)}+\Phi_g^{(1)}]\otimes
H^{(0)}$
($H^{(0)}\otimes[\Phi_d^{(1)}+\Phi_e^{(1)}+\Phi_f^{(1)}+\Phi_g^{(1)}]$).

The infrared-finite $k_T$-dependent NLO hard kernel for
Fig.~\ref{leading}(a) is given by
\begin{eqnarray}
H^{(1)}
& = & { \alpha_s(\mu_{\rm f}) C_F \over 4 \pi } \bigg( {21 \over 4}
\ln { \mu^2 \over Q^2} -6 \ln { \mu_{\rm f}^2 \over Q^2}
 - {17 \over 4}\ln^2 x_1  + {27 \over 8} \ln x_1 \ln x_2
-\ln^2 \delta_{12}+{17 \over 4} \ln x_1 \ln \delta_{12}\nonumber
\\
&&  - {13 \over 8} \ln x_1 + {31 \over 16} \ln x_2 + {23 \over 8}
\ln \delta_{12} + {1 \over 2} \ln 2+ { 5 \pi^2 \over 48} + {53
\over 4} \bigg)H^{(0)}. \label{pht}
\end{eqnarray}
We have chosen $|n^+|= n^-$, which renders $\zeta^2= Q^2$, to avoid
creating an additional large logarithm like $\ln(\zeta^2/Q^2)$. The
Sudakov logarithm $\ln^2(Q^2/k_T^2)$ in Eq.~(\ref{double}) has been
removed by that in the effective diagrams, but the threshold
logarithms $\ln^2x^2$ remains in $H^{(1)}$, which can be absorbed
into a jet function
\cite{UL}. Because there is no end-point singularity in the pion
form factor at twist-2, we shall not perform the factorization of this
threshold logarithm here. However, an end-point singularity does
exist in the contribution from the two-parton twist-3 pion distribution
amplitudes, so the factorization
and the threshold resummation for the jet function needs to be
performed.

We have
checked that Eq.~(\ref{pht}) is gauge-invariant by repeating
the NLO calculations in an arbitrary covariant gauge. Applying
the Ward identity to the gauge-dependent terms \cite{NL07,LM09},
it is easy to show that the gauge dependence cancels between the
quark diagrams and the effective diagrams.
The NLO hard kernel ${\bar H}^{(1)}$ for Fig.~\ref{leading}(b) can
be obtained from Eq.~(\ref{pht}) by exchanging $x_1$ and $x_2$,
and by substituting ${\bar H}^{(0)}$ in Eq.~(\ref{hbar}) for $H^{(0)}$.
The NLO hard kernels for the other two LO quark
diagrams with the virtual photon attaching to the anti-quark line
can be obtained from $H^{(1)}$ and ${\bar H}^{(1)}$ by exchanging
$x$ and $1-x$.

\section{NUMERICAL ANALYSIS}

In this section we evaluate the NLO correction to the pion
electromagnetic form factor in the $k_T$ factorization theorem numerically.
Since the Sudakov suppression was derived in the impact parameter
space, we write the factorization formula as a convolution in
the impact parameters $b_1$ and
$b_2$ \cite{LS}, conjugate to $k_{1T}$ and $k_{2T}$,
respectively. The Sudakov factor, resulting from the summation
of the double logarithm $\alpha_s\ln^2k_T$ to all orders, describes
the perturbative $k_T$ dependence of a pion wave function. We shall not
consider the intrinsic $k_T$ dependence of a pion wave function
proposed in \cite{JK93} here. To simplify the expression of the NLO
hard kernel in the $b_1$-$b_2$ space, we adopt the
approximation $\ln\delta_{12}\approx \ln(x_1x_2)$. This
approximation does not modify the
power-law behavior of end-point contributions from small $x$,
and affects numerical outcomes only slightly. We
first test the following asymptotic models for the twist-2 pion distribution
amplitude $\phi_\pi^A$ and the two-parton twist-3 pion distribution
amplitudes $\phi_\pi^{P,T}$,
\begin{eqnarray}
 \phi_{\pi}^A(x) &=& \frac{6f_{\pi}}{2\sqrt{2N_c}} x(1-x), \nonumber\\
 \phi_{\pi}^P(x) &=& \frac{f_{\pi}}{2\sqrt{2N_c}}, \nonumber\\
 \phi_{\pi}^T(x) &=& \frac{f_{\pi}}{2\sqrt{2N_c}}(1-2x),\label{as}
\end{eqnarray}
where the pion decay constant is taken  as $f_{\pi} = 130$ MeV.

\begin{figure}[t]
\begin{center}
\includegraphics[height=7cm]{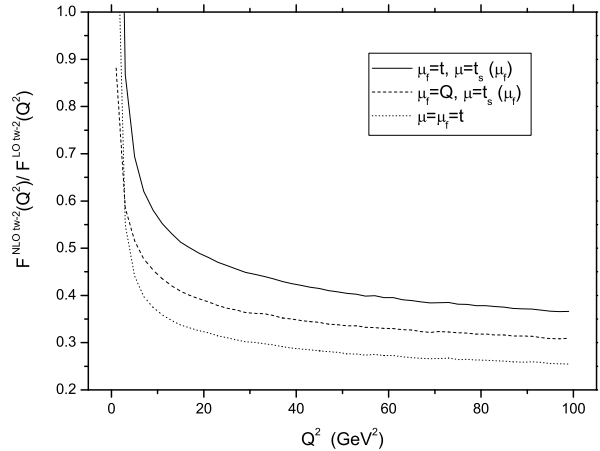}
\caption{Ratio of the NLO correction over the LO contribution to
the pion form factor with the scale $t_s (\mu_{\rm f})$ being
defined in Eq. (\ref{ts function}). } \label{ratio}
\end{center}
\end{figure}

The first issue concerns the choice of the renormalization scale
$\mu$ and the factorization scale $\mu_{\rm f}$ in order to minimize
the NLO correction to the pion form factor. The results are summarized in
Fig.~\ref{ratio}, which displays the ratio of the NLO contribution over
the LO one as a function of $Q^2$. For the first choice,
$\mu_{\rm f}$ is set to the hard scale $t$ defined in the PQCD approach to
exclusive processes based on the $k_T$ factorization theorem \cite{LS,KLS}
\begin{eqnarray}
t=\max(\sqrt{x}Q,1/b_1,1/b_2).
\end{eqnarray}
Then we utilize the
freedom of choosing $\mu$ to diminish all the single-logarithmic
and constant terms in the NLO hard kernel, which is found to be
\begin{eqnarray}
t_s (\mu_{\rm f})  = \exp\left({53 \over 42}+{5 \pi^2 \over
504}\right) x_1^{5/42} x_2^{11/24} \left({\mu_{\rm f} \over
Q}\right)^{1/7} \mu_{\rm f}. \label{ts function}
\end{eqnarray}
This scale choice gives a larger NLO correction:
the NLO correction becomes less than
40\% of the LO contribution only when $Q^2$ is higher than 55
GeV$^2$. The second choice corresponds to $\mu_{\rm f}=Q$, which
is greater than $t$, and $\mu=t_s$, which removes all the
single-logarithmic and constant terms in the NLO hard kernel.
The convergence of the NLO
correction is improved: the NLO correction becomes
less than 40\% as $Q^2>17$ GeV$^2$. The third choice is the
conventional one adopted in the PQCD approach to exclusive
processes, with both $\mu$ and $\mu_{\rm f}$ being set to $t$.
It turns out that this simple choice works better:
the NLO correction becomes less than 40\% as $Q^2 > 7$ GeV$^2$.
The above analysis confirms the prescription postulated in
\cite{LS} that both the
renormalization and factorization scales are set to the invariant masses
of internal particles.

\begin{figure}[t]
\begin{center}
\includegraphics[height=7cm]{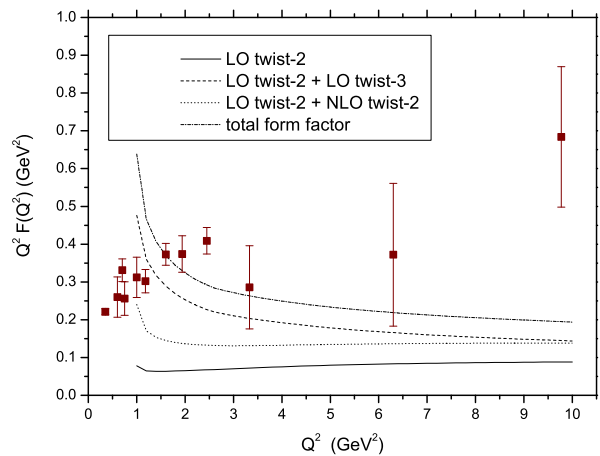}
\caption{Pion form factor from different orders and twists for the
asymptotic pion distribution amplitudes in Eq.~(\ref{as}). The
experimental data are taken from \cite{Huber:2008id} and
\cite{Bebek:1977pe}.} \label{asy}
\end{center}
\end{figure}

After fixing the scales, we calculate the LO twist-2, NLO twist-2,
and LO two-parton twist-3 contributions to the pion form factor,
and present the results in Fig.~\ref{asy}. Note that the last
piece of contribution involves two two-parton twist-3 pion
distribution amplitudes, so it is suppressed by a power of
$1/Q^2$. Figure~\ref{asy} indicates that the LO twist-2
contribution is quite small compared to the experimental data in
the whole considered range of $Q^2$. The NLO twist-2 contribution
is also small, and helps to explain the data only at $Q^2\sim 1$
GeV$^2$, where the perturbation theory may not be reliable: the
quick growth of our results at very low $Q^2$ is attributed to the
presence of the infrared Landau pole in the strong coupling
constant. To study the pion form factor in this region, a soft
contribution is needed, and an analytic perturbation theory may
help \cite{BPSS}. The LO two-parton twist-3 contribution dominates
in most range of $Q^2$, and becomes comparable to the NLO twist-2
one as $Q^2\sim 10$ GeV$^2$. The dominance comes from the chiral
enhancement scale $m_0({\rm 1 GeV}) = 1.74$ GeV, which is of the
same order of magnitude as low $Q$ values. With this contribution
being included, it is possible to accommodate the data in the
large $Q^2$ region. It supports the treatment in the PQCD approach
to exclusive $B$ meson decays \cite{KLS}, in which the NLO twist-2
correction to $B$ transition form factors is neglected, but the LO
two-parton twist-3 one is included. That is, the latter should be
the most important subleading contribution. The recent BaBar data
for the pion transition form factor at low $Q^2$ favors an
asymptotic pion distribution amplitude \cite{BABAR}.
Figure~\ref{asy} implies that the data of the pion form factor at
low $Q^2$ also support the asymptotic models for the pion
distribution amplitudes. Besides, the asymptotic models are
favored from the viewpoint of light-cone sum rules, which
reproduce the data of the pion form factor \cite{Braun:1999uj}. As
pointed out in \cite{Blok:2002ew}, the experimental data of
$F(Q^2)$ at higher $Q^2$ from \cite{Bebek:1977pe} have
large additional uncertainties due to the model dependent
substraction of transverse cross section, on top of the
large statistical and systematic errors. Therefore, no conclusive
statement on the agreement or disagreement between data and theory
can be made.

\begin{figure}[t]
\begin{center}
\includegraphics[height=7cm]{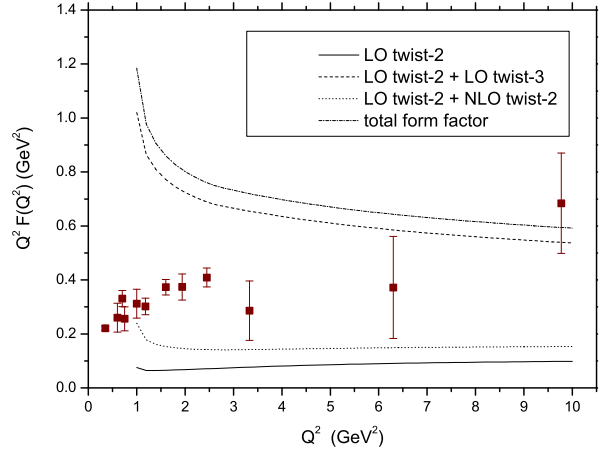}
\caption{Contributions to the pion form factor from different orders and twists
for the non-asymptotic models in Eq.~(\ref{non}).} \label{pion}
\end{center}
\end{figure}

It is interesting to examine whether the shape of the pion
distribution amplitudes matters for explaining the data, an issue
that has been investigated and discussed
in various aspects intensively. We test the following
non-asymptotic pion distribution
amplitudes \cite{Duplancic:2008ix}:
\begin{eqnarray}
 \phi_{\pi}^A(x) &=& \frac{6f_{\pi}}{2\sqrt{2N_c}} x(1-x)[ 1 +0.16 C_2^{3/2}(u)+0.04 C_4^{3/2}(u)], \nonumber\\
 \phi_{\pi}^P(x) &=& \frac{f_{\pi}}{2\sqrt{2N_c}}[1 +0.59 C_2^{1/2}(u) +0.09 C_4^{1/2}(u) ],
 \nonumber\\
 \phi_{\pi}^T(x) &=& \frac{f_{\pi}}{2\sqrt{2N_c}}[C_1^{1/2} (u)+0.019 C_3^{1/2}
 (u)],\label{non}
\end{eqnarray}
with the Gegenbauer polynomials
\begin{eqnarray}
 &C^{1/2}_{1}(u)=u ,&C_3^{1/2} (u) = \frac{1}{2} u (5u^2 -3),  \nonumber\\
 &C_2^{1/2}(u)=\frac{1}{2} (3u^2-1),& C_4^{1/2} (u)=\frac{1}{8}
(35 u^4- 30 u^2 +3),\nonumber \\
 % C^{1/2}_{3}(t)=\frac{1}{2}({5t^{3}}-{3t}), &C_{3}^{3/2}(t)=\frac{1}{2}({35 t^3}-{15t}) ,
 %\\
&C_2^{3/2} (u)=\frac{3}{2} (5u^2-1), & C_4^{3/2} (u)=\frac{15}{8}
(21 u^4 -14 u^2 +1),
\end{eqnarray}
and the variable $u=1-2x$. Figure~\ref{pion} shows that the above
models overshoot the data in the small  $Q^2$ region. Hence, the
matrix elements of higher conformal spin operators may have been
overestimated in \cite{Duplancic:2008ix}.

\begin{figure}[t]
\begin{center}
\includegraphics[height=7cm]{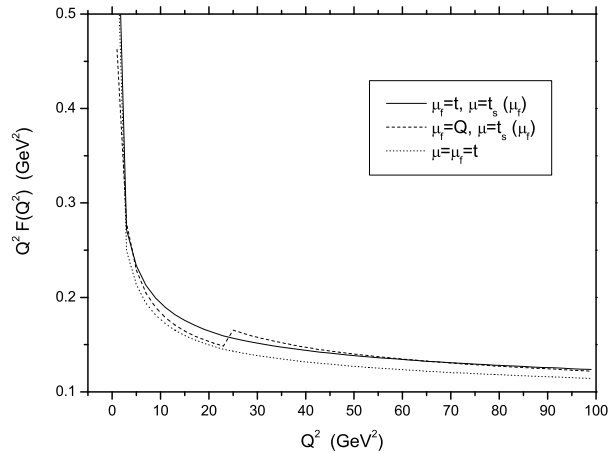}
\caption{Scale dependence of the pion form factor with
$t_s (\mu_{\rm f})$ being defined in Eq. (\ref{ts
function}). } \label{QQ}
\end{center}
\end{figure}

At last, we investigate the theoretical uncertainly in our calculation
by studying the scale dependence, and the results are presented in
Fig.~\ref{QQ}. The curves corresponding to $\mu_{\rm f}=t$, $\mu=t_s$
and $\mu_{\rm f}=\mu=t$ differ by about 10\% in
the considered range of $Q^2$, implying that
the theoretical uncertainty is under control. The sharp jump
exhibited by the curve corresponding to $\mu_{\rm f}=Q$, $\mu=t_s$ is
attributed to the change of the QCD scale $\Lambda_{\rm QCD}$ with the
upper bounds of the integration variables $b_1$ and $b_2$. When the
momentum transfer squared increases up to $Q^2=23$ GeV$^2$,
the value $\Lambda_{\rm QCD}^{(4)} =286$ MeV is
replaced by $\Lambda_{\rm QCD}^{(5)} =204$ MeV, and
the sharp jump is produced as shown in Fig.~\ref{QQ}. This jump is not
obvious for $\mu_{\rm f}=t$ (the other two curves), whose value is
typically smaller than $Q$, so it does not excite $\Lambda_{\rm
QCD}^{(4)}$ to $\Lambda_{\rm QCD}^{(5)}$ at $Q^2=23$ GeV$^2$.

\section{CONCLUSION}

In this paper we have extended the framework for higher-order
calculations in the $k_T$ factorization theorem to the pion
electromagnetic form factor. The
prescription is that partons in both QCD quark diagrams and
effective diagrams for TMD hadron wave functions are off mass shell
by $k_T^2$. The light-cone divergences
in naive definitions of TMD hadron wave functions were
regularized by rotating the Wilson lines away from the light
cone. Since the $k_T$ factorization theorem is
appropriate for QCD processes dominated by contributions from small
momentum fractions, terms suppressed by powers of momentum fractions
have been neglected, and simple expressions can be obtained for the box
and pentagon diagrams. We have demonstrated the disappearance of the soft
logarithms, and the exact cancellation of the
collinear logarithms in the difference of the quark and
effective diagrams. Our work represents the first NLO
analysis of the pion form factor in the $k_T$ factorization theorem.

Our numerical study has indicated that setting the
renormalization and factorization scales to invariant masses of
internal particles reduces NLO corrections. The NLO correction to
the pion form factor becomes lower than 40\% of
the LO one for $Q^2>7$ GeV$^2$. This observation
supports the conventional scale choice adopted in the PQCD approach to
exclusive processes based on the $k_T$ factorization theorem, and
is consistent with that made in the collinear factorization
theorem \cite{MNP99}.  We have found that the NLO
twist-2 contribution does not help accommodating the
data of the pion form factor. Instead, the LO two-parton twist-3
contribution plays a crucial role for the purpose. It is also confirmed that the
asymptotic pion distribution amplitudes are favored by the data
available at low $Q^2$. Comparing the results from the
different choices of the renormalization and factorization
scales, the theoretical uncertainty in our calculation is
estimated to be around 10\%. We shall apply the same framework to
the NLO analysis of $B$ meson transition form factors in a
forthcoming paper.

\vskip 0.3cm  The work was supported in part by the National Science
Council of R.O.C. under Grant No. NSC-98-2112-M-001-015-MY3, by the
National Center for Theoretical Sciences of R.O.C., by National
Science Foundation of China under Grant No. 11005100, and by
the the Deutsche Forschungsgemeinschaft under Contract No.
KH205/1-2. Y.M.W. would like to acknowledge Prof. Cai-Dian L\"{u}
for warm hospitality during his visit at IHEP, Beijing, and Prof.
Xue-Qian Li for inviting him to give a talk at Nankai University.

\end{document}